\documentclass[runningheads]{llncs}
\usepackage{graphicx}
\usepackage{multirow}
\usepackage{amsmath}
\usepackage{booktabs}
\usepackage{makecell}
\usepackage{graphicx}
\usepackage{enumerate}
\usepackage{enumitem}

\begin{document}
\title{Stock Volatility Prediction Based on Transformer Model Using Mixed-Frequency Data}
\titlerunning{Prediction Based on Transformer and Mixed-Frequency Data}
%
\author{Wenting Liu\inst{1} \and Zhaozhong Gui\inst{2} \and Guilin Jiang\inst{3}\thanks{Corresponding author: Guilin Jiang, jiangguilin@hnchasing.com.} \and Lihua Tang\inst{2} \and Lichun Zhou\inst{1} \and Wan Leng\inst{2} \and Xulong Zhang\inst{4} \and Yujiang Liu\inst{5} }

\authorrunning{Wenting Liu and Zhaozhong Gui and et al.}
%
\institute{Chasing Jixiang Life Insurance Co., Ltd.\and Hunan Chasing Digital Technology Co., Ltd. \and Hunan Chasing Financial Holdings Co., Ltd. \and  Ping An Technology (Shenzhen) Co., Ltd. \and The University of Melbourne}
%
\maketitle              
\begin{abstract}
With the increasing volume of high-frequency data in the information age, both challenges and opportunities arise in the prediction of stock volatility. On one hand, the outcome of prediction using tradition method combining stock technical and macroeconomic indicators still leaves room for improvement; on the other hand, macroeconomic indicators and peoples’ search record on those search engines affecting their interested topics will intuitively have an impact on the stock volatility. For the convenience of assessment of the influence of these indicators, macroeconomic indicators and stock technical indicators are then grouped into objective factors, while Baidu search indices implying people’s interested topics are defined as subjective factors. To align different frequency data, we introduce GARCH-MIDAS model. After mixing all the above data, we then feed them into Transformer model as part of the training data. 
Our experiments show that this model outperforms the baselines in terms of mean square error. The adaption of both types of data under Transformer model significantly reduces the mean square error from 1.00 to 0.86.

\end{abstract}
\section{Introduction}

Measuring and predicting market risk is the primary prerequisite for managing and controlling financial markets. Among them, the volatility of financial assets is a commonly used characteristic indicator to measure the risk in them, making it also a core issue in financial research. 
However, the volatility of financial assets cannot be directly observed through our eyes. 
To solve this problem, extensive research has been conducted on measurement methods for volatility. In the early days, volatility was directly measured by variance or standard deviation. Various models were constructed to evaluate volatility, such as ARCH (AutoRegressive Conditional Heteroskedasticity) model~\cite{engle1982autoregressive} and SV (Stochastic Volatility) model~\cite{taylor1994modeling}. These models are the basic models for studying financial time series and can reflect the fluctuation characteristics of variances. On this basis, the focus of research has gradually shifted to predicting the volatility of financial assets.

With the development of machine learning techniques, models like LM(Long Memory)~\cite{Gilles2004Volatility} and Markov-switching model~\cite{2009Estimating} were introduced subsequently and significantly reduced the prediction error compared to traditional statistic models. Machine unlearning methodology was optimized on the basis of Stochastic Teacher Network~\cite{zhang2023Machine}.
Indicators such as the popularity of daily news and investors' sentiment were also incorporated into models~\cite{古红宏2020引入投资者情绪的沪深}.

However, current research faced a unified problem in selecting auxiliary indicators for volatility prediction. Firstly, the data of macroeconomic variables is usually produced on a monthly basis, while that of financial assets is by minute or even by second. So the difference in data frequency to consider various types of indicators in volatility prediction. Secondly, investors' subjective emotions greatly affect their investment behavior. Therefore, how to choose indicators that can well reflect investors' subjective emotions is also a challenge. 
In order to address this issue, GARCH-MIDAS (Generalized AutoRegressive Conditional Heteroskedasticity and Mixed Data Sampling) model~ \cite{engle2008spline}  was proposed for data processing, to extract macroeconomic information so as to incorporate more objective factors reflecting volatility changes. 

Recently, volatility prediction models have been extended to deep learning models~\cite{2019Performance}. Models such as LSTM(Long Short-Term Memory)~\cite{2018Forecasting}, TabNet~\cite{0Stock} and CNN(Convolutional Neural Network)~\cite{2020Goldcnn} were introduced into this area and all demonstrated outstanding performance in mixing different types of data and predicting volatility.
At the same time, self-attention-based architectures, in particular the Transformers, have become the up-to-date method in more and more fields such as NLP(Natural Language Processing)~\cite{2017Attention} and CV(Computer Vision)~\cite{2020vit}. Motivated by their breakthroughs, we introduce the Transformer model to the prediction of financial market.

The outputs of the GARCH-MIDAS model are deployed to train the Transformer model, which is one of the innovations of this paper. This paper also selects Baidu search index as as an indicator of investors' sentiment. It is derived from the search activity of Baidu users for specific keywords on the Baidu search engine.
By integrating these metrics, we observe a notable enhancement in the predictive capabilities of our Transformer model.
Our contributions are summarized as follows:
\begin{enumerate}[label=\arabic*)]
    \item To enrich data dimensions by incorporating macroeconomic and investor attention factors. We applied macroeconomic data information of different frequencies to volatility prediction and used multiple Baidu indices to measure investor attention, thereby improving the effectiveness of volatility prediction.
    \item To evaluate the applicability and prediction effectiveness of deep learning techniques for stock-related data. Given the various models applied to volatility prediction, there is still room for enhancement. We used the Transformer model to effectively improve the prediction accuracy of the model, thereby demonstrating the effectiveness of this model in predicting volatility.
\end{enumerate}

\section{Related Works}

With the progress and development of economy and society, the interest and depth of research on financial asset volatility are increasing day by day. Recent studies on volatility can be mainly divided into 2 categories from a research perspective:


The first category focuses on studying volatility from a prediction perspective, exploring based on different types of data and models. Choudhury et al. \cite{2014A} used support vector machines to predict future prices and developing short-term trading strategies based on the predictions, and the test simulation achieved good profits within 15 days. S. Chen et al. \cite{陈声利2018基于跳跃} established a HAR volatility modeling framework based on the Baidu search index, incorporating it with the jumping, good and bad volatility optimization models. They then evaluated the effectiveness of the model through MCS testing. In the work by Hu \cite{2020A}, a novel hybrid approach was introduced for forecasting fluctuations in copper prices. This innovative method synergistically integrates the GARCH model, Long Short-Term Memory (LSTM) network, and conventional Artificial Neural Network (ANN), yielding commendable accuracy in predicting price volatility. 
Y. H. Umar and M. Adeoye
 \cite{umar2020markov} estimated the volatility using the Markov regime conversion method by comparing all monthly stock index data of the Central Bank of Nigeria (CBN) from 1988 to 2018 in the statistical bulletin of the Nigerian Stock Exchange. B. Schulte-Tillman \cite{2022Financial} proposed four multiplicative component volatility MIDAS models to distinguish short-term and long-term volatility, and found that specific long-term variables in the MIDAS model significantly improved prediction accuracy, as well as the superior performance of a Markov switching MIDAS specification (in a set of competitive models). A. Vidal et al. \cite{2020Goldcnn} used a CNN-LSTM model to predict gold volatility. At the same time in deep learning field, A. Vaswani et al. \cite{2017Attention} and A. Dosovitskie et al. \cite{2020vit} proposed to use Transformer to replace LSTM and CNN in NLP and CV field.

The second category focuses on studying the factors that influence volatility, emphasizing on analyzing the factors that affect volatility and their impact. C. Christiansen et al. \cite{2010A} conducted an in-depth investigation into the drivers behind fluctuations in financial market volatility. Their study encompassed a thorough exploration of the predictive influence exerted by macroeconomic and financial indicators on market volatility. R. Hisano et al. \cite{hisano2013high} undertook an assessment of the influence of news on trading dynamics. Their analysis encompassed an extensive dataset of over 24 million news records sourced from Thomson Reuters, examining their correlations with trading behaviors within the S\&P US Index's prominent 206 stocks.
C. A. Hartwell \cite{hartwell2018impact} constructed a unique monthly database from 1991 to 2017 to explore the impact of institutional fluctuations on financial volatility in transition economies. F. Audrino et al. \cite{audrino2020impact} adopted a latest sentiment classification technique, combining social media, news releases, information consumption, and search engine data to analyze the impact of emotions and attention variables on stock market volatility; F. Liu et al. \cite{刘凤根2020基于混频数据模型的宏观经济对股票市场波动的长期动态影响研究} studied the long-term dynamic situation of volatility from two levels: horizontal values and volatility, and selected four macroeconomic variables to analyze their impact. P. Wang's \cite{Peng2020Impact} study concentrated on important stocks in the stock exchange market's financial sector. The objective was to assess the influence of margin trading and stock lending on the price volatility of these chosen stocks. The findings revealed a noteworthy observation: both margin trading activities and the balances associated with such trading had the potential to amplify the level of volatility in stock prices.

\section{Methodology}
In this section, a briefing of the volatility theory and feature extraction method will be given. How the GARCH-MIDAS and the Transformer are deployed in the prediction of stock volatility will also be explained.

\subsection{Basic Theory of Volatility}
In order to explore the real market volatility , 
this article uses RV (Realized Volatility) as an indicator to measure the volatility of the CSI300. 
The calculation of RV was defined by Andersen and Bollerslev (1998) \cite{1998Answering}, with the following specific formulas:

\begin{equation}
    R_{t}=100(\ln Pr_t-\ln Pr_{t-1}) 
\end{equation}
\begin{equation}
    R_{t,d}=100(\ln Pr_{t,d}-\ln Pr_{t,d-1} ) 
\end{equation}
\begin{equation}
    RV_{t}=\sum_{d=1}^{48} R_{t,d}^2 
\end{equation}
where $Pr_t$, $R_t$ and $RV_t$ represent the price, the return and RV on the t-day, respectively.
 $Pr_{t,d}$ and $R_{t,d}$ represent the price and the return on the 5-minute interval
 of the t-day, respectively.

However, it is well known that the trading of stock market occurs within a limited
time rather than 24 hours without interruption. Hansen and Lunde (2005) \cite{2005A} further demonstrated that the previously defined RV lacks information during non-trading time and proposed 
to use scale parameter to adjust it appropriately. The approach proposed by these scholars is written into the adjusted formula below. 
\begin{equation}
    \lambda = \frac{\sum_{t=1}^N R_t^2 /N}{\sum_{t=1}^N RV_t / N} 
\end{equation}
\begin{equation}
    RV_t' = \lambda \times RV_t
\end{equation}
where $\lambda $ is the scale parameter and  $ RV_t' $ stands for the adjusted RV on the t-day.
    
\subsection{Definition of Indicators}
The research object of this article is CSI300, which integrates 
the information of the top 300 excellent stocks. 
The frequency for calculating RV is every 5 minutes. In addition, the data of Baidu search index is obtained by crawling from the internet using python software.
In order to measure 
the impact of various factors on stock returns, we refer to the research of 
scholars S. Li \cite{李树阳2019基于循环神经网络及百度指数预测} and M. Zhang \cite{张苗苗0基于}. And we determine the final indicators according to the grey correlation degree of 
indicators and returns, as well as the Baidu demand map. The objective factors and subjective factors are as shown in Table~\ref{tab1}.

\begin{table}[t]
\caption{Meaning of Volatility Related Factors}\label{tab1}
\resizebox{\linewidth}{!}{
\begin{tabular}{c|ccc}
    \toprule
    \multicolumn{2}{c}{Factor} 
    & Index & Variable \\
    \midrule
\multirow{22}{*}{\makecell[c]{Objectivity \\ Factors}} &
\multirow{10}{*}{\makecell[c]{ Macroeconomic \\ Indicators}} 
& Macro-economic Consensus Index (Current Value)  & MeCI \\
& & Macro-economic Leading Index (Current Value) & MeLeI \\
& & Macro-economic Lagging Index (Current Value) & MeLaI \\
& & Consumer Price Index (CPI, Last month = 100) & CPI \\
& & Total Retail Sales of Consumer Goods (Current Value/Yuan) & Retailsale \\
& & Retail Price Index (RPI, Last month = 100) & RPI \\
& & Producer Price Index (PPI, Last month = 100) & PPI \\
& & Money Supply (Total Balance at End of Period, Yuan) & M2 \\
& & Fixed Asset Investment (Cumulative Value, Yuan) & FInvest \\
& & Total Imports and Exports (Current Value/US Dollars) & IOP \\
\cmidrule{2-4}
& \multirow{12}{*}{\makecell[c]{Stock \\ Technical \\ Indicators}}
& Turnover Rate(\%) & Turn \\
& & Bollinger Bands Indicator 
(Median Line/Number of Periods 26) & BOLL \\
& & 5-day Moving Average & MA(5) \\
& & 20-day Moving Average & MA(20) \\
& & Moving Average Convergence Divergence & MACD \\
& & Relative Strength Index (Number of Periods 6) & RSI \\
& & Selling On-Balance Volume & SOBV \\
& & Rate of Change & ROC \\
& & Trading Volume & Volume \\
& & Highest Price & High \\
& & Lowest Price & Low \\
& & Open Price & Open \\
\midrule
\multirow{5}{*}{\makecell[c]{Subjective \\ Factors}} &
\multirow{5}{*}{\makecell[c]{Attention \\ Indicators}} &
``CSI300'' Baidu Search Index & CSI300 \\
& & ``CSI500'' Baidu Search Index & CSI500 \\
& & ``SSE50'' Baidu Search Index & SSE50 \\
& & ``Components of CSI300'' Baidu Search Index & HSparts \\
& & ``CSI300 Index Fund''  Baidu Search Index & HSETF \\
\bottomrule

\end{tabular}
}
\end{table}

Considering the large number of indicators, the inconsistent scales among them and the 
characteristics of data (such as non-stationarity, large fluctuations, 
and missing values), pre-processing of all data is required before 
model construction for later use and analysis. The pre-processing 
mainly includes two aspects: missing value filling and data normalization. 
In addition, in order to minimize information loss of each indicator and avoid multiple collinearity between indicators, we use PCA(Principal Component Analysis) methods to extract 
principal components and construct a comprehensive index.

In terms of macroeconomic indicators, two principal components (PCM1 and PCM2) are extracted, 
which together capture 98.3\% of the information. PCM1 carries a major positive load distribution on indices e.g. CPI, Retailsale and 
Finvest, so it is labeled 
as ``consumption and investment component''. PCM2 carries
a major positive load on indices e.g. RPI, PPI, etc., so it is labeled as ``production and prosperity component''. In terms of stock technical indicators, we extract three principal 
components (TECH1, TECH2 and TECH3), with a total contribution ratio close to 100\%. 
TECH1 consists of indices like 
average price, highest price and lowest price, mainly reflecting the size of CSI300, so it is 
labeled as ``price component''. TECH2 
consists of indices like ROC, MACD 
and etc., reflecting the stock price changes and market 
attention of CSI300, so it is labeled as ``trend component''. TECH3 consists of 
indices like Turn and Volumn, reflecting the liquidity of CSI300, 
so it is labeled as ``liquidity component''. Based on the results of 
the scree plot, we get the investor attention component (BD1), with a variance contribution rate of 
approximately 88.6\%, which was labeled as the ``attention component''. The detailed compositions of all the above principal components are recorded in Table 2.

\begin{table}
\centering
\caption{The load value of the principal components of each indicator.}\label{tab2}
\begin{tabular}{crr|crrr|cr}
\hline
\multicolumn{3}{c}{Macroeconomic Indicators} & 
\multicolumn{4}{c}{Stock Technical Indicators} & 
\multicolumn{2}{c}{Attention Indicators} \\
 & PCM1 & PCM2 & & TECH1 & TECH2 & TECH3 & & BD1 \\ 
\hline
MeCI & 0.24 & 0.5 & Turn & 0.17 & 0.21 & 0.96 & CSI300 & 0.77 \\
MeLeI & 0.02 & -0.14 & BOLL & 0.97 & -0.04 & 0.21 & CSI500 & 0.55 \\
MeLaI & 0.15 & 0.16 & MA5 & 0.97 & 0.08 & 0.23 & SSE50 & 0.32 \\
CPI & 0.89 & -0.07 & MA20 & 0.97 & -0.02 & 0.22 & HSparts & 0.03 \\
Retailsale & 0.94 & -0.07 & MACD & 0.1 & 0.75 & 0.26 & HSETF & 0.05 \\
RPI & -0.06 & 0.85 & RSI & 0.05 & 0.88 & 0.09 &  &  \\
PPI & 0.24 & 0.75 & SOBV & 0.9 & 0.05 & -0.14 &  &  \\
M2 & -0.37 & -0.12 & ROC & -0.01 & 0.94 & 0.06 &  &  \\
FInvest & 0.49 & -0.48 & Volume & 0.33 & 0.2 & 0.91 &  &  \\
IOP & 0.06 & -0.47 & High & 0.96 & 0.11 & 0.24 &  &  \\
 &  &  & Low & 0.97 & 0.12 & 0.21 &  &  \\
 &  &  & Open & 0.96 & 0.11 & 0.23 &  &  \\

\hline
\end{tabular}
\end{table}

\subsection{Prediction Method}
\begin{figure}
    \centering
    \includegraphics[width=\textwidth]{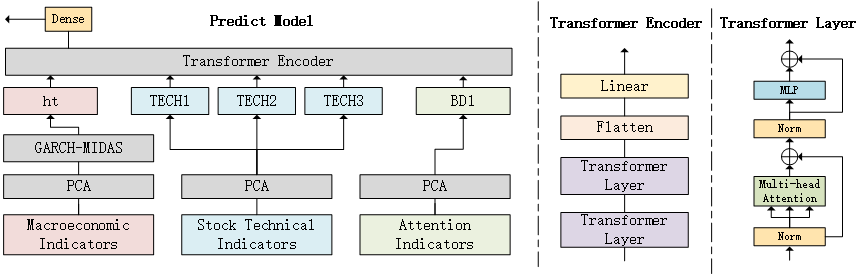}
    \caption{The Structure Diagram of the Transformer Network} \label{fig3}
\end{figure}

As discussed in Session 3.2, we reserve only the principal components of the related factors, align the frequencies and feed them into Transformer model. The prediction method of this article, as shown in Figure~\ref{fig3}, mainly includes the following three steps:
\begin{enumerate}[label=\arabic*)]
    \item Extracting principal components for each factor. Macroeconomic indicators, stock technical indicators, and subjective factors are sequentially extracted using PCA method to obtain six principal components (PCM1, PCM2, TECH1, TECH2, TECH3 and BD1).
    \item Training the mixed-frequency data model, which feeds the daily returns of the two principal components of macroeconomic indicators (PCM1, PCM2) and CSI300 into the GARCH-MIDAS model (Session 3.3.1), ultimately obtains the conditional volatility ht.
    \item Using the Transformer model (Session 3.3.2) to train the conditional volatility (ht), taking the principal components of stock technical indicators (TECH1, TECH2, TECH3), and the principal components of Baidu index as input variables and obtaining the prediction results.
\end{enumerate}

\subsubsection{GARCH-MIDAS Model}
\ 
\newline
In the analysis of stock volatility, using monthly or quarterly data to construct models will lose high-frequency effective information of stock market. Therefore, Ghysels et al. \cite{Eric2006Predicting} first proposed the Mixed Sampling Model (MIDAS), and Engle et al. \cite{2008The} further applied this model to the Generalized Autoregressive Conditional Heteroscedasticity Model, forming the GARCH-MIDAS model. The GARCH-MIDAS model's return and volatility are described as follows:
\begin{equation}\label{eq9}
    r_{i,t}-E_{i-1,t}\left(r_{i,t}\right)=\sqrt{\tau_tg_{i,t}}\varepsilon_{i,t}, \forall i=1,2,...,N_{t}
\end{equation}
\begin{equation}
    \varepsilon_{i,t}|\psi_{i-1,t}\sim N(0,1) \notag
\end{equation}
\begin{equation}\label{eq10}
    \sigma_{i,t}^2=E\left[\left(r_{i,t}-E_{i-1,t}\left(r_{i,t}\right)\right)^2\right]=\tau_tg_{i,t}
\end{equation}
$N_{t}$ represents the number of days in t-th month. $r_{i,t}$, $\psi_{i,t}$ and $g_{i,t}$ correspond to the return, the information set of the yield and  the high-frequency fluctuations on the i-th day of the t-th month, respectively. And $E_{i-1,t}\left(r_{i,t}\right)$ represents the 
conditional mathematical expectation under the condition when the information 
set $\psi_{i,t}$ is given at the (i-1)-th moment of time with the market
return $r_{i,t}$. $\tau_t$ reflects the low-frequency 
fluctuations in the t-th month, and $\sigma_{i,t}^2$ is the conditional variance.

Assuming that the conditional mathematical expectation of $r_{i,t}$ at the (i-1)-th 
moment is $\mu$ and that the short-term component of the returns follows a 
$GARCH(1,1)$ process, Formula~(\ref{eq9}) can be rewritten as Formula~(\ref{eq11}), with 
short-term fluctuations given by Formula~(\ref{eq12}). 
At this point, long-term 
fluctuations are represented by the filtering equation for realized 
volatility, which is given by Formula~(\ref{eq13}). In this equation, $\theta$ represents
the long-term component indicates the 
contribution of volatility to its marginal. $RV_{t-k}$ is the volatility 
of market returns over a fixed time horizon, and $\phi_k(\omega_1,\omega_2)$ is the 
weight function equation, and K is the maximum lagging order of the low-frequency.
\begin{equation}\label{eq11}
    r_{i,t}=\ \mu+\sqrt{\tau_tg_{i,t}}\varepsilon_{i,t}
\end{equation}
\begin{equation}\label{eq12}
    g_{i,t}=\omega+\frac{{\alpha\left(r_{i,t}-\mu\right)}^2}{\tau_t}+\beta g_{i-1,t}
\end{equation}
\begin{equation}\label{eq13}
    \tau_t=m+\theta\sum_{k=1}^{K}{\varphi_k(\omega_1,\omega_2){RV}_{t-k}}
\end{equation}
\begin{equation}\label{eq14}
    \varphi_j\left(\omega_1,\omega_2\right)=\frac{{(k/K)}^{\omega_1-1}\times{(1-k/K)}^{\omega_2-1}}{\sum_{k=1}^{K}{{(k/K)}^{\omega_1-1}\times{(1-k/K)}^{\omega_2-1}}}
\end{equation}

In addition, $m$, $\mu$, $\omega_1$, $\omega_2$ and $\theta$ are all parameters to be estimated. Generally, $\omega_1$ is fixed to 1 to confirm that the weight of the lagged variable exhibits a decaying trend. $\omega_2$ reflects the decay rate of the impact of the low-frequency on the high-frequency.

\subsubsection{Transformer Model}
\ 
\newline
The Transformer model was raised by Vaswani et al~\cite{2017Attention}, a deep learning model on the basis of self-attention mechanisms. The core idea of the Transformer model is to treat each element in the input sequence as a vector and use self-attention mechanisms to compute the relationships between these vectors. Practically, the attention function is as follows:
\begin{equation}
Attention(Q, K, V) = softmax(\frac{QK^T}{\sqrt{d_k}})V
\end{equation}
where $Q$, $K$, and $V$ are the abbreviation of 'Query', 'Key', and 'Value' respectively. Each representing an element in the input sequence, which is usually a vector. A dot product operation is applied to calculate the similarity between $Q$ and $K$, and then use the softmax function to convert it into a probability distribution. ${\sqrt{d_k}}$ is a scaling factor, helps the model better capture dependencies in the input sequence and improves its performance.
Multi-head attention is a mechanism used in the Transformer model to compute the relationships between different positions in the input sequence. It is based on the idea of applying self-attention to each element in the input sequence separately, but with different weights for each attention head.

\begin{align}
MultiHead(Q,K,V)  &=  Concate(head_1, \dots, head_h)W^O \\
head_i  &= Attention(QW_i^Q, KW_i^K, VW_i^V) \notag
\end{align}
$W_i^Q  \in R^{d_o \times d_q} $, 
$W_i^K  \in R^{d_o \times d_k}$, 
$W_i^V \in R^{d_o \times d_v} $ 
and $ W^O  \in R^{hd_v \times d_o} $.

Traditionally, the Transformer model consists of multiple encoders and decoders stacked together. Whereas, when dealing with problems in CV, it is suggested to only feed the resulting sequence to one Transformer encoder, and introduce a Multi-Layer Perceptron (MLP) to form a classification or regression head. ~\cite{2020vit}

As we have already extracted the principal components and reduced the dimensions, we can simply concatenate data from consecutive days and treat it as independent data. Therefore, we follow the ViT model closely in model design, except for we do not have to do positional encoding.

\section{Empirical analysis}
\subsection{Experiment Setup}
To prove the validity of model construction, 
the prepared data is split into training and testing data with 
a 9:1 ratio. An Transformer model using a python software 
package named keras is developed, which then shows excellent long-term memory ability  for financial time series in the process of continuous input data streaming.
\subsubsection{ Data Reparation}
\ 
\newline
The data to be fed into the GARCH-MIDAS model contains macroeconomic indicators and returns of CSI300, which are collected from Jan 2011 to Sep 2021 on a monthly and daily basis, respectively. The GARCH-MIDAS model is then implemented using this data and the fit\_mfgarch function of R software. The optimal value of the lag period $K$ of the GARCH-MIDAS model is decided to be 12 after repeated trials. The output results are recorded in Table~\ref{tab3}.

\begin{table}
\centering
\caption{Estimated values of main parameters for the GARCH-MIDAS model. (Note: * and ** show significance at 5\%, and 1\% levels, separately.)}\label{tab3}
\begin{tabular}{crrcrr}
\toprule 
\makebox[0.1\textwidth][c]{Parameter} & \makebox[0.15\textwidth][c]{\makecell[c]{Estimated \\ Value}} & \makebox[0.2\textwidth][c]{P-value} & \makebox[0.1\textwidth][c]{Parameter} & \makebox[0.15\textwidth][c]{\makecell[c]{Estimated \\ Value}} & \makebox[0.2\textwidth][c]{P-value} \\
\midrule
$\mu$ & 0.046755 & 0.037720*\,\,\, & $\theta^{(1)}$ & -0.376158 & 0.025157*\,\,\, \\
$\alpha$ & 0.071928 & 0.000002** & $\omega^{(1)}_2$ & 63.666123 & 0.000000** \\
$\beta$ & 0.911217 & 0.000000** & $\theta^{(2)}$ & -0.760231 & 0.020216*\,\,\, \\
$m$ & 0.730420 & 0.012188*\,\,\, & $\omega^{(2)}_2$ & 1.395697 & 0.000000** \\
\bottomrule
\multicolumn{6}{l}

\end{tabular}
\end{table}

It can be observed from Table~\ref{tab3} that the sum value of parameters $\alpha$ and $\beta$ is close 
to 1, which indicates a well-fit for the short-term fluctuations of CSI300, and a convergence of the conditional variance of the model to the mean at a appropriate speed. 

Parameter $\theta^{(1)}$ represents the ``consumption and investment component'', with a negative value of about -0.376, indicating a high volatility of CSI300 when the 
consumption and investment values are small. Generally, the decline of consumption levels implies a decrement of people's willingness and ability to invest. People tend to be more conservative and cautious, which may significantly influence the stock prices. Parameter $ \theta^{(2)} $ corresponds to the ``production 
and prosperity component'', with a negative value of about -0.760, indicating a high volatility of CSI300 when the production and prosperity levels are low. It is commonly understood that a decrement of production 
may cause a supply-demand imbalance and poor circulation 
of the market. 

Parameter $\omega^{(1)}_2$ is the weight of  $\theta^{(1)}$, while parameter $\omega^{(2)}_2$ is the weight of $ \theta^{(2)} $. A lower value of  $\omega^{(2)}_2$ respective to $\omega^{(1)}_2$ indicates a lower dependency of the model on ``production 
and prosperity components'' as compared to ``consumption and investment component''.

\begin{figure}
    \includegraphics[width=\textwidth]{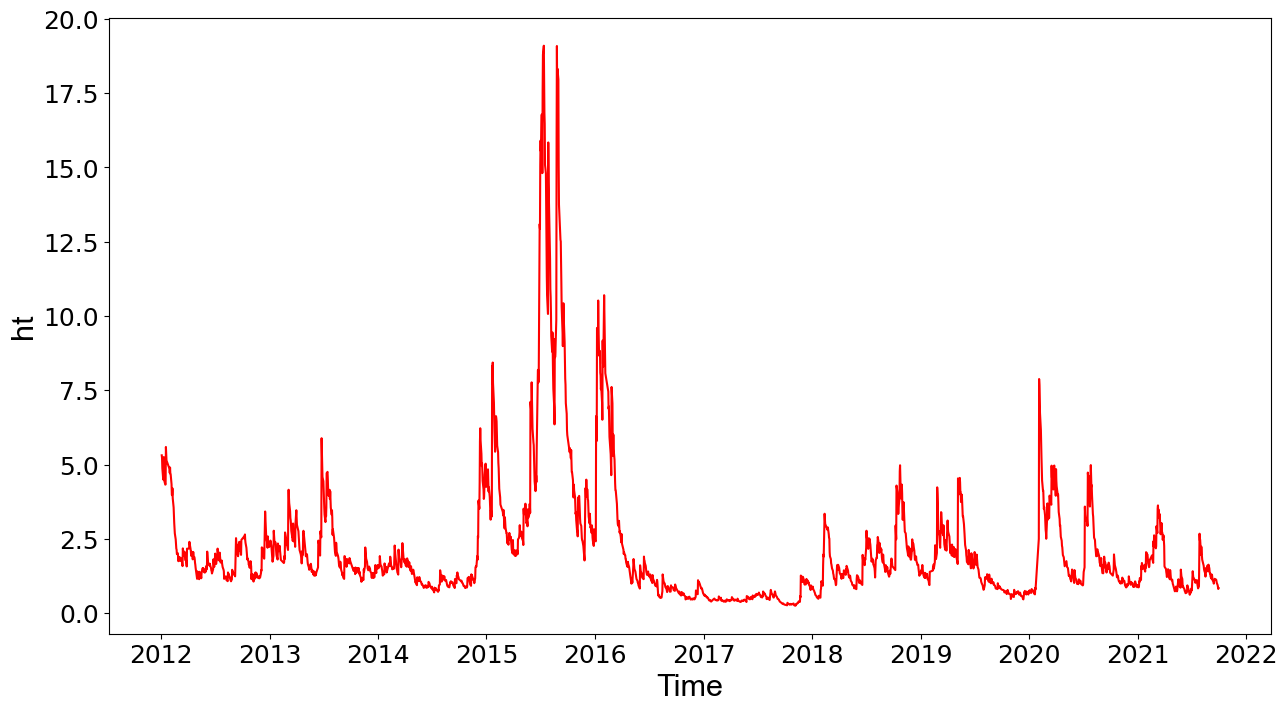}
    \caption{Estimation of Conditional Volatility(ht)} \label{fig2}
\end{figure}

The conditional volatility of the model is shown in Figure~\ref{fig2}. Due to the lag period setting of 12 months, the parameter estimation period will start from 2012. The conditional volatility includes information from macroeconomic indicators, thereupon alleviates other factors' impact on volatility. In 2015, the conditional volatility of CSI300 was intense, indicating that it experienced remarkable ups and downs during this period, which was closely related to the crash of stock market during that year. In firstly half of 2015, China's macro-control over real estate gradually strengthened, leading to further warming of the investment market and the increment of investors' enthusiasm. However, from mid-June of 2015, the stock market experienced a sharp decline in stock prices and significant fluctuations. CSI300 was also greatly affected. In the end of 2019 and beginning of 2020, the sudden outbreak of COVID-19 in Wuhan led to nationwide shutdowns, resulting in crucial impacts on the stock market and causing fluctuations of CSI300. This figure proves that the conditional volatility of GARCH-MIDAS model can well reflect the actual situation and the effectiveness of the model.
\begin{table}
\centering
\caption{Values of Hyperparameters in the Transformer Model.}\label{tab_hyper}
\begin{tabular}{ccl}
\toprule
Parameter   & \multicolumn{1}{c}{Value} & \multicolumn{1}{c}{Remarks}                          \\
\midrule
TimeStepSize & 5 & The number of days of data used in the prediction process. \\
LearningRate & 0.05   & The magnitude of weight updates of each round.       \\
BatchSize    & 32     & The number of data samples that are passed to model. \\
NumHeads     & 3      & The number of heads of Multihead Attention.          \\
NumLayers    & 2      & The number of transformer layers.       \\
\bottomrule
\end{tabular}
\end{table}

\subsubsection{Hyperparameter Setting} 
\ 
\newline
The Transformer model involves many hyperparameters that require multiple attempts to find the optimal state. The main hyperparameters to be used by the filters of the model are interpreted in Table~\ref{tab_hyper}.

\subsection{Experiment Result}

\subsubsection{The Prediction Result of the Transformer Model}
\begin{figure}
    \includegraphics[width=\textwidth]{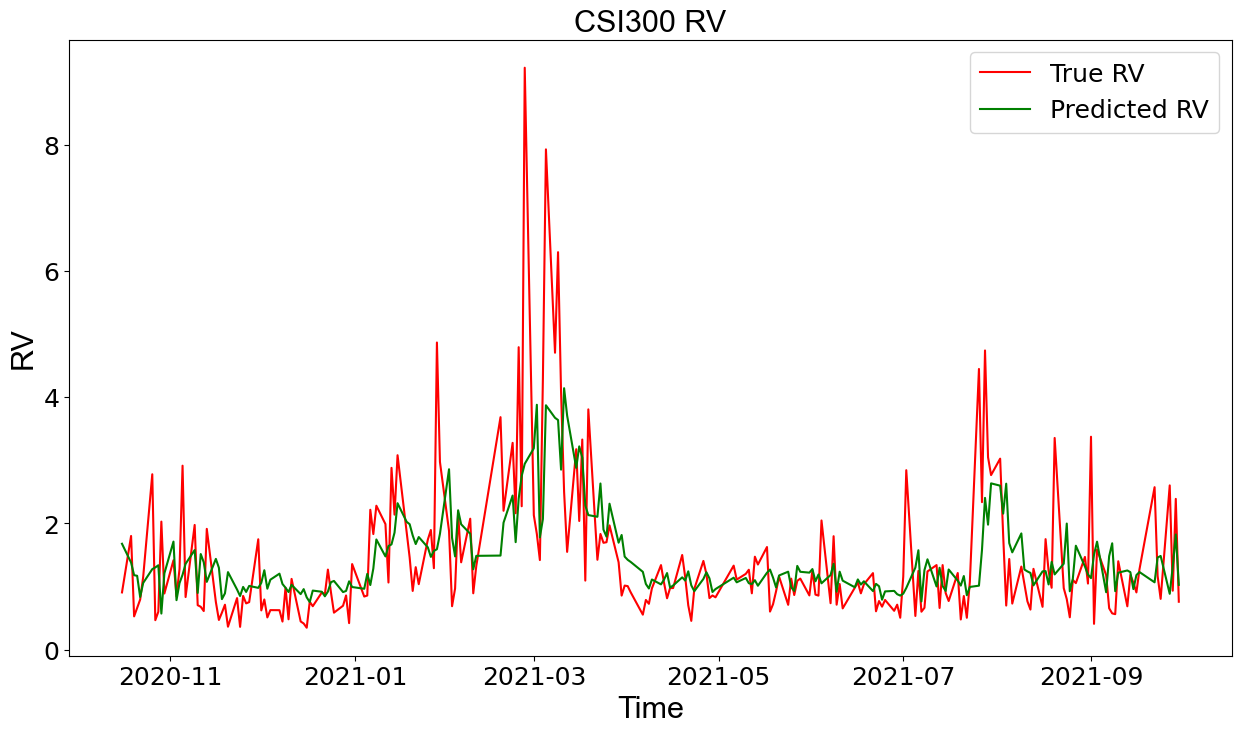}
    \caption{Prediction Result of the Transformer Model} \label{fig4}
\end{figure}

\ 
\newline
The volatility from Oct 2020 to Sep 2021 is generated using the model tuned during training. The predicted RV is compared with the true RV in Figure~\ref{fig4}.

Figure~\ref{fig4} shows the predictive results of the Transformer model on the basis of GARCH-MIDAS and PCA on the test set. The red curve represents 
the RV size of CSI300 daily, 
while the green curve reflects the predicted volatility generated by the Transformer 
model. Overall, the model gives a good evaluation in some turning points 
and rising trends.
\subsubsection{Comparison of Different Factors}
\ 
\newline

The changes in the macroeconomic environment will have an impact on 
factors such as capital costs and discount rates. Investors' attention 
is an important factor that affects investment behavior and can 
lead to the stock market turbulence. Therefore, this article aims to verify 
the importance of these two factors by grouping different types of factors 
from the training data for comparison.
\begin{itemize}
    \item Group G1: Stock technical indicators only; 
    \item Group G2: Attention indicators and stock technical indicators; 
    \item Group G3: Macroeconomic indicators and stock technical indicators; 
    \item Group G4: Macroeconomic indicators, Attention indicators and stock technical indicators.

\end{itemize}

\begin{table}
    \centering
    \caption{Prediction Accuracy Assessment of the Transformer Model with Different Indicator Groups.}\label{tab4}
    \begin{tabular}{crrrrrr}
    \toprule 

    Indicator Group &  MSE & HMSE &  MAE & MAPE & QLIKE & $\rm R^2LOG$ \\
    \midrule
    G1 & 0.9951 & 0.6067 & 0.6317 & 0.5402 & 1.3860 & 0.1474 \\
    G2 & 0.8973 & 0.4720 & 0.6016 & 0.4905 & 1.3620 & 0.1626 \\
    G3 & 0.9666 & 0.4843 & 0.6136 & 0.5082 & 1.3425 & 0.1266 \\
    G4 & 0.8624 & 0.4460 & 0.5871 & 0.4787 & 1.3620 & 0.1710 \\
    
    \bottomrule
    \end{tabular}
\end{table}
Table~\ref{tab4} includes the evaluation results of 6 loss functions for 
different models. The results of G4 compared with the rest of the 
models show that the overall MSE, HMSE, MAE, MAPE and QLIKE results are 
smaller than the other 3 groups, which indicates that incorporating 
macroeconomic indicators and subjective attention can improve the accuracy 
of model predictions, and also indicates that these two types of indicator factors have an 
impact on the fluctuation of CSI300. When 
comparing G2 and G3 with G1 individually, it is found 
that adding either macroeconomic indicators or subjective attention only will also increase 
the prediction accuracy effectively. Therefore, when predicting the 
fluctuation rate of CSI300, incorporating 
both macroeconomic indicators and subjective attention as input features 
has a good synergy effect, and the improvement in model prediction 
accuracy is more obvious.

\subsubsection{Comparison of Different Models}
\ 
\newline
In terms of stock market forecasting, many scholars 
have attempted various methods and continuously improved their prediction 
accuracy. Among those commonly used deep learning models, we choose 4 of them to compare with our Transformer model, i.e. LSTM, CNN, XGBoost and GRU(Gate Recurrent Unit).
\begin{table}
    \centering
    \caption{Prediction Accuracy of Different Models.}\label{tab5}


\begin{tabular}{lrrrrrr}
\toprule
\multicolumn{1}{c}{Model} &
  \multicolumn{1}{c}{MSE} &
  \multicolumn{1}{c}{HMSE} &
  \multicolumn{1}{c}{MAE} &
  \multicolumn{1}{c}{MAPE} &
  \multicolumn{1}{c}{QLIKE} &
  \multicolumn{1}{c}{R2LOG} \\
\midrule
Transformer & 0.8624 & 0.4460 & 0.5871 & 0.4787 & 1.3620 & 0.1710 \\
LSTM        & 0.8801 & 0.6689  & 0.6496 & 0.6008 & 1.4014 & 0.0671 \\
CNN         & 1.4716 & 1.1846 & 0.8466 & 0.7631 & 1.5496 & 0.0720 \\
XGBoost     & 0.9421 & 0.4963 & 0.6030  & 0.4936 & 1.3343 & 0.1322 \\
GRU         & 1.5563 & 1.8534 & 0.9844 & 1.0391 & 1.6627 & 0.0317 \\
\bottomrule
\end{tabular}
\end{table}

According to the results in Table~\ref{tab5}, the Transformer model performs better than 
others commonly used for volatility prediction. It has the best 
results in terms of MSE, HMSE and MAPE loss functions, and its overall 
prediction accuracy is good. This indicates that the Transformer model can 
effectively extract the characteristics of CSI300 
volatility, and is more suitable for predicting this volatility as compared 
to other models.

\section{Conclusion}
This article addresses the problem of combination use of different frequencies between macroeconomic data and daily stock data. On the basis of GARCH-MIDAS model, the monthly information from macroeconomic indicators is converted to daily information as an input feature for the later Transformer model. The parameters of the GARCH-MIDAS model are remarkable, demonstrating that the converted daily information can well include macroeconomic information. In terms of selecting macroeconomic indicators, ten representative indicators are finally selected through grey correlation analysis to eliminate the influence of subjective selection and information redundancy. We would like to provide a new insight for future research on the application of mixed-frequency data in predicting volatility of financial assets.

This paper Takes objective and subjective factors as input features of the Transformer network, determining the main parameters of the model through empirical experiments. The adjusted RV is used as an alternative to reflect the real volatility and evaluate the validation of the Transformer model. In addition, the effects and accuracy of the GARCH-MIDAS and PCA models are analyzed from both a factor and a model perspective. The results show that the addition of macroeconomic indicators and attention indicators can increase the predictive accuracy of the transformer model, and the transformer model has a advantage over other models in predicting CSI300 volatility. The results of the Transformer model are also ideal, and the loss functions are within a reasonable range.

\bibliographystyle{splncs04}
\bibliography{main}

\end{document}